\documentclass[print,twocolumn]{revtex4}
\usepackage[dvips]{graphicx}
\usepackage{graphicx}
\usepackage{dcolumn}
\usepackage{bm}
\usepackage{epsfig}
\usepackage{latexsym,bm}

\begin{document}

\title{The applications of the general and reduced Yangian algebras}
\author{Li-Guo Qin}
\author{Li-Jun Tian}\email{tianlijun@staff.shu.edu.cn}
\author{Yan-Ling Jin}
\author{Guo-Hong Yang}
\affiliation{Department of Physics, Shanghai University,
Shanghai,
200444, China}

\begin{abstract}
The applications of the general and reduced Yangian $Y(sl(2))$ and
$Y(su(3))$ algebras are discussed. By taking a special constraint,
the representation of $Y(sl(2))$ and $Y(su(3))$ can be divided into
two $2$ $\times$ $2$ and three $3$ $\times$ $3$ blocks diagonal
respectively. The general and reduced Yangian $Y(sl(2))$ and
$Y(su(3))$ are applied to the bi-qubit system and the mixed light
pseudoscalar meson state, respectively. We can find that the general
ones are not able to make the initial states disentangled by acting
on the initial states, however the reduced ones are able to  make
the initial state disentangled. In addition, we show the effects of
$Y(su(3))$ generators on the the decay channel.\\

PACS number(s): 02.20.-a, 03.65.-w, 13.25.-k


\end{abstract}

\maketitle \baselineskip=12pt

\section{Introduction}

Quantum entanglement, a nonlocal correlation, is of crucial
importance in quantum computation \cite{Nielsen}, quantum
teleportation \cite{Bennett1}, dense coding \cite{Bennett2} and
quantum key distribution \cite{Curty}. However, in a real system,
the deterioration of the coherence or even the decoherence and
disentanglement due to the interaction with its surroundings, which
is recognized as a main obstacle to realize quantum computing
\cite{Beige} and quantum information processing (QIP) \cite{Viola},
have to be taken into account in the researches in the field of
quantum information. Earlier studies had indicated that entanglement
decays exponentially \cite{Yu1, Yu2, Privman} until T. Yu suggested
that entanglement decays completely in finite time and called for
concerted effect to research entanglement sudden death \cite{Yu}.
For example, different systems \cite{Yonac, Ikram, Cao} and
realizations in experiment\cite{Almeida, Rau} provide theoretical
guidance to practical application of controlling entanglement.

In the last decades, Yangian algebras associated with simple Lie
algebras has been studied systematically in both mathematics and
physics, and have many applications through spin operators and
quantum fields \cite{3bernard, 3Uflov, 3Inozemtsev}. There have been
some remarkable successes in studying the long-ranged interaction
models by Yangian approaches \cite{ap, fd1, fd2, ml1, mw, ml2} in
which the Haldane-Shastry model was regarded as the representative
of the spin chain $su(n)$ with long-range interaction \cite{fd2}.
Recently, Yangian $Y(sl(2))$ and $Y(su(3))$ have been studied for
quantum entanglement \cite{ljtq,ljtq2}. In addition, Yangian
$Y(su(3))$ algebra has been demonstrated to be able to realize the
hadronic decay channels of light pseudoscalar mesons and predict a
possible explanation of the unknown particle $X$ in the decay
channel $K_L^0\rightarrow\pi^0\pi^0X$ \cite{Tian1}. In this paper,
the influence of transition operators composed of the generators of
Yangian $Y(sl(2))$ and $Y(su(3))$ on the entanglement degrees of
two-qubit system and the mixed light pseudoscalar meson states are
discussed in the cases of the general and reduced Yangian algebras
respectively.

In this work, we will study the effects of the generators of Yangian
$Y(sl(2))$ on the entanglement degrees of two-qubit system in two
cases: the general and reduced Yangian. Then we will use the similar
method in \cite{Tian4} to make the $Y(su(3))$ algebra reduced by
taking a special constraint, namely, the $Y(su(3))$ algebra becomes
the block-diagonal form. Finally, we would like to discuss the
applications of the general and reduced Yangian $Y(su(3))$ algebras
to the mixed light pseudoscalar meson states. And also, some
examples are presented to compare the effect of the transition
operators of the general Yangian algebras with reduced ones on the
entanglement degrees. Results show that the generators of the
reduced Yangian algebras can make the final states disentangled,
while the general ones can not make it disentangled under the same
condition.

The paper is organized as follows. In Sec. II, the applications of
the general and reduced Yangian $Y(sl(2))$ algebra in the Bi-qubit
system are discussed, and the main results are presented. In Sec.
III, The reduced Yangian $Y(su(3))$ algebras is obtained by taking a
special constraint. Sec. IV presents the applications of the general
and reduced Yangian $Y(su(3))$ Algebra in the mixed light
pseudoscalar meson states. In Sec. V, conclusions are presented.

\section{The applications of Yangian $Y(sl(2))$ Algebra in the Bi-qubit System }

To compare the effects of the general and the reduced $Y(sl(2))$ on the Bi-qubit system,
we will firstly reduce the general $Y(sl(2))$ to the reduced one,
then show the effects of them on the entanglement of the Bi-qubit system.

The Yangian $Y(sl(2))$ is generated by the generators
$\{{I}_{\alpha},{J}_{\alpha}\}$ with the commutation relation
\cite{Ge4}:
\begin{eqnarray}
[{I}_\alpha,{I}_\beta]=i\epsilon_{\alpha\beta\gamma}{I}_\gamma,\;\;\;\;
[{I}_\alpha,{J}_\beta]=i\epsilon_{\alpha\beta\gamma}{J}_\gamma
\;\;\;\;(\alpha,\beta,\gamma=1,2,3),\nonumber
\end{eqnarray}
where the $\{I_{\alpha}\}$ form a simple Lie algebra $ sl(2) $
characterized by $\epsilon_{\alpha\beta\gamma}$ and
\begin{eqnarray}
&&[{J}_{\pm},[{J}_3,{J}_\pm]]=\frac{h^2}{4}{I}_{\pm}
({J}_{\pm}{I}_3-{I}_{\pm}{J}_3),\nonumber
\end{eqnarray}
where $h$ is the deformation parameter and the notations
${I}_\pm={I_1}{\pm}i{I_2}$ and ${J}_\pm={J_1}{\pm}i{J_2}.$

Now let us consider a bi-spin system, the realization of generators
of $Y(sl(2))$ take the form of \cite{Ge4}
\begin{eqnarray}
{\bf{\emph{\textbf{I}}}}={\bf{\emph{\textbf{S}}}}={\bf{\emph{\textbf{S}}}_1}+{\bf{\textbf{S}}_2},
\end{eqnarray}
\begin{eqnarray}
{\bf{\emph{\textbf{J}}}}=\frac{\mu}{\mu+\nu}{\bf{\emph{\textbf{S}}}}_1}+\frac{\nu}{\mu+\nu}{\bf{\emph{\textbf{S}}_2}+\frac{i\lambda}{\mu+\nu}{\bf{\emph{\textbf{S}}_1}\times{\bf{\emph{\textbf{S}}}}_2},
\end{eqnarray}
where ${\bf{\emph{\textbf{S}}_1}}$, ${\bf{\emph{\textbf{S}}_2}}$ are
the spin-$\frac{1}{2}$ operators and $\mu$, $\nu$ and $\lambda$ are
arbitrary parameters. ${\bf{\emph{\textbf{I}}}}$ is the total spin
operator satisfying $[I_i^a,I_j^b]=i \epsilon_{abc} I_i^c
{\delta}_{i j},(i, j=1, 2)$.

With a special constraint relation $\mu\nu=-\frac{1}{4}\lambda^{2}$
\cite{Tian4}, we can get $J^2=\frac{3}{4}$,
$[J_{a},J_{b}]=i\epsilon_{abc}J_c$. Similarity transformations of
the generators can be made by the use of the matrix $\tau$ who takes
the form of
\begin{eqnarray}
\label{A1} &&\tau=\left(\matrix{1&0&0&0\cr
0&\nu&-{\frac{1}{2}}{\lambda}&0\cr 0&
-{\frac{1}{2}}{\lambda}&\nu&0\cr 0&0&0&1}\right).
\end{eqnarray}
After the similarity transformations, the generators become
\begin{eqnarray}
\label{y+1}
&&Y^+=\tau^{-1}J^{+}\tau=\left(\matrix{\frac{\xi{\sigma}^{+}}{2}&0\cr0&\frac{\xi^{-1}{\sigma}^{+}}{2}}\right),\nonumber\\
&&Y^-=\tau^{-1}J^{-}\tau=\left(\matrix{\frac{\xi^{-1}{\sigma}^{-}}{2}&0\cr 0&\frac{\xi{\sigma}^{-}}{2}}\right),\nonumber\\
&&Y^3=\tau^{-1}J^{3}\tau=\frac{1}{2}\left(\matrix{\sigma^3&0\cr \cr 0&\sigma^3}\right),\nonumber\\
\end{eqnarray}
where $\xi$=${\nu}-\frac{1}{2}{\lambda}$ and $\sigma$ are pauli
matrices. \{$Y^{a}, a=\pm, 3$\} reduce to two $2\times2$ blocks
diagonal and $4\times4$ matrix is essentially the 4-dimension
representation of $sl(2)$ algebra, so it is marked as the reduced
$Y(sl(2))$ algebra in this case.

We will study the effects of the $Y(sl(2))$ algebra generators on
the entanglement degree of Bi-qubit system as follows.

For an arbitrary two-qubit pure state
$|\Phi\rangle=a_{00}|00\rangle+a_{01}|01\rangle+a_{10}|10\rangle+a_{11}|11\rangle$,
where $a_{00}$, $a_{01}$, $a_{10}$, and $a_{11}$ are the normalized
complex amplitudes, the concurrence (the measurement of
entanglement) C is given by \cite{F}
\begin{eqnarray}
\label{6} C=2|a_{00}a_{11}-a_{01}a_{10}|\;\; and \; \; 0\leq C\leq1
.
\end{eqnarray}
with the maximally entangled state (MES) $C=1$ and the separable
state $C=0$. Of course, we can construct another general state as
the initial state
\begin{eqnarray}
|\phi\rangle=\frac{1}{\sqrt2}[\alpha(|00\rangle+|11\rangle)+\beta(|01\rangle+|10\rangle)]
\end{eqnarray}
where $|\alpha|^2+|\beta|^2=1$.

The concurrence C of the initial state is
\begin{eqnarray}
\label{c}
 C=|\alpha^2-\beta^2|.
\end{eqnarray}
By making use of the transition characteristic of Yangian operators
$J^{i}$ and $Y^{i}$ ($i=+,-,3$) \cite{ljtq,ljt}, we can construct
transition operators $P$, which can make the initial state transit
another state
\begin{eqnarray}
\label{p} |\phi_{i}\rangle=P|\phi\rangle.
\end{eqnarray}
To illustrate the relation of every transition operator versus the
concurrence, we will discuss two cases as follows.

(i) The general $Y(sl(2))$

Due to the transition effect of Yangian generators, transition
operators $P$ can be constructed as compositions of Yangian
$Y(sl(2))$ generators.

(a) Let us first take the transition operator
\begin{eqnarray}
P_{1} = J^+,
\end{eqnarray}
which is composed of the general Yangian $Y(sl(2))$ generators. By
acting $P_{1}$ to the initial state, we can get the final state
$|\phi_{1}\rangle=P_{1}
|\phi\rangle=\frac{1}{\sqrt2}(\frac{\nu+\frac{\lambda}{2}}{\mu+\nu}\alpha|01\rangle+
\frac{\mu-\frac{\lambda}{2}}{\mu+\nu}\alpha|10\rangle+\beta|11\rangle)$
with the normalization condition
\begin{eqnarray}
|\frac{\nu+\frac{\lambda}{2}}{\mu+\nu}\alpha|^2 +
|\frac{\mu-\frac{\lambda}{2}}{\mu+\nu}\alpha|^2+ |\beta|^2=2.
\end{eqnarray}

The concurrence of the final state $|\phi_{1}\rangle$ is obtained as
\begin{eqnarray}
\label{c1}
C_{1}=|\frac{(\mu-\frac{\lambda}{2})(\nu+\frac{\lambda}{2})}{2(\mu+\nu)^2}\alpha^2|,
\end{eqnarray}

Taking the transition operator as $P_{2}=J^{-}$, we can obtain
$|\phi_{2}\rangle=P_{2}
|\phi\rangle=\frac{1}{\sqrt2}(\beta|00\rangle+\frac{\mu-\frac{\lambda}{2}}{\mu+\nu}\alpha|01\rangle
+\frac{\nu+\frac{\lambda}{2}}{\mu+\nu}\alpha|10\rangle)$. The
concurrence of $|\phi_{2}\rangle$ is same with $\phi_{1}$, namely,
$C_1=C_2$ with range from $0$ to $1$.

(b)Choose the transition operator
\begin{eqnarray}
P_{3} = J^3,
\end{eqnarray}
we can obtain the final state
$|\phi_{3}\rangle=\frac{1}{\sqrt2}[\alpha(-|00\rangle+|11\rangle)-\frac{\mu-\nu-\lambda}{\mu+\nu}\beta(|01\rangle-|10\rangle)]$.
Utilizing the normalization condition, the concurrence is gotten as
\begin{eqnarray}
 C_3=C=|\alpha^2-\beta^2|,
\end{eqnarray}
which is the same with the concurrence of the initial state in Eq.
(\ref{c}). That is to say, the transition operator $J^{3}$ can not
change the entanglement degree of the initial state.

(ii) The reduced $Y(sl(2))$

(a) When we take the transition operator
\begin{eqnarray}
P_{4} = Y^+,
\end{eqnarray}
which is composed of the reduced Yangian $Y(sl(2))$ generators. By
acting $P_{4}$ to the initial state, we will obtain the final state
$|\phi_{4}\rangle=\frac{1}{\sqrt2}(\xi^{-1}\alpha|01\rangle+\xi\beta|11\rangle)$.
The corresponding normalization condition leads to
\begin{eqnarray}
|\xi^{-1}\alpha|^2 + |\xi\beta|^2=2.
\end{eqnarray}

The concurrence of the final state $|\phi_{4}\rangle$ is
\begin{eqnarray}
\label{c3} C_{4}=0,
\end{eqnarray}
When $P_5=Y^{-}$, the final state yields
$|\phi_{5}\rangle=\frac{1}{\sqrt2}(\xi\beta|00\rangle+\xi^{-1}\alpha|10\rangle)$.
We can obtain the same concurrence of the final state
$|\phi_{5}\rangle$ with $C_{4}$. So we can obtain that the reduced
Yangian $Y^{\pm}$ can make the initial state disentangled.

(b) If we choose the transition operator
\begin{eqnarray}
P_{6} = Y^3
\end{eqnarray}
corresponding to the final state
$|\phi_{6}\rangle=\frac{1}{\sqrt2}[\alpha(-|00\rangle+|11\rangle)+\beta(|01\rangle-|10\rangle)]$.
Utilizing the normalization condition, the concurrence is gotten as
\begin{eqnarray}
 C_4=C=|\alpha^2-\beta^2|,
\end{eqnarray}
which is the same with the concurrence of the initial state in Eq.
(\ref{c}). That is to say, the transition operator $Y^{3}$ can not
change the entanglement degree of the initial state too. By
comparing the effects of the transition operators of the general and
reduced Yangian on entanglement, we find that the reduced Yangian
operators ($Y^+$ and $Y^-$)make the initial state disentangled
directly, but the general Yangian operators can not. In addition,
through carefully observing, we find that $\alpha$ is the parameter
of $|00\rangle$ and $|11\rangle$ at first, but become the parameter
of $|01\rangle$ and $|10\rangle$ after action of $P_1$, $P_2$, $P_4$
and $P_5$, namely, these four operators can make the transformation
happen between states. However, there is not so transformation for
$P_{3}$ and $P_6$. $J^+$ and $Y^+$ make $\beta$ transition to
$|11\rangle$, not $|00\rangle$. However, the same case, $J^-$ and
$Y^-$ make $\beta$ transition to $|00\rangle$, not $|11\rangle$. By
above different transition cases, we can fully understand the
transition effect of Yangian $Y(sl(2))$ operators.

\section{The Reduced $Y(su(3))$ Algebra}

The subalgebra of $Y(su(3))$ is Lie algebra $su(3)$, which we have
been familiar with the $su(3)$ symmetry of elementary particles
\cite{3GellMann, 3Ne'eman}. $su(3)$ generators are defined by
$[F^{a},F^{b}]=if_{abc}F^{c}$ ( $a$, $b$, $c$ $=$ 1, 2, $\cdots$,
8), where the structure constants $f_{abc}$ are antisymmetric
$f_{123}=1,
f_{458}=f_{678}=\frac{\sqrt{3}}{2},f_{147}=f_{246}=f_{257}=f_{345}=-f_{156}=-f_{367}=\frac{1}{2}$.
The 3-dimensional representation of $su(3)$ is formed by the
well-known Gell-Mann matrices, i.e.,
\begin{eqnarray}
\label{f}&&{\Lambda}^a=2F^a,\;\;\;\{\Lambda^a,a=1,2,\cdots 8 \}\nonumber\\
&&[\Lambda^{a},\Lambda^{b}]=2if_{abc}\Lambda^{c}.
\end{eqnarray}

Now we can introduce the shift operators
\begin{eqnarray}
\label{+-} &&I^{\pm}=F^{1}{\pm}iF^{2},\;\;\;
U^{\pm}=F^{6}{\pm}iF^{7},\;\;\;\;\;
V^{\pm}=F^{4}{\mp}iF^{5},\nonumber\\
&& I^{3}=F^{3},\;\;\;\;\;\;\;\;\;\;\;
Y=\frac{2}{\sqrt{3}}F^{8}=I^{8},\;\;(Y-hypercharge)\nonumber\\
\end{eqnarray}

and the notations
\begin{eqnarray}
\label{u3v3} &&U^3=-\frac{1}{2}I^{3}+\frac{3}{4}I^8,\;\;\;
 V^3=-\frac{1}{2}I^{3}-\frac{3}{4}I^8,\nonumber\\
 &&Q=I^3+\frac {1}{2}Y.
\end{eqnarray}
where $\vec{U}$ (${U^i, i=\pm, 3}$), $\vec{V}$ (${V^i, i=\pm, 3}$)
and $Q$ are called the $U$-spin, $V$-spin and the charge operator
respectively. And it is easy to check that $[U^i,Q]=0$, but there is
no the same property between $V$-spin and $I$-spin.

The fundamental representation of local $su(3)$ is given by
\begin{equation}
\label{iuv38}
\begin{array}{l}
\begin{array}{lll}
\vspace{0.3cm} I^{+}=\left( \matrix{0&1&0 \cr 0&0&0 \cr
0&0&0}\right)&\;\;\; U^{+}=\left( \matrix{0&0&0 \cr 0&0&1 \cr
0&0&0}\right) \cr V^{+}=\left( \matrix{0&0&0 \cr 0&0&0 \cr
1&0&0}\right)&\;\;\; I^{3}=\frac{1}{2}\left( \matrix{1&0&0 \cr
0&-1&0 \cr \vspace{0.3cm}
 0&0&0}\right) \cr
 Y=\frac{1}{3}\left( \matrix{1&0&0 \cr 0&1&0
\cr 0&0&-2}\right),
\end{array}
\end{array}
\end{equation}
and $I^{-}$=$(I^{+})^{+}$, $U^{-}$=$(U^{+})^{+}$,
$V^{-}=(V^{+})^{+}$.

  For the system of two particles, we can define the operators of $Y(su(3))$ as follows:
\begin{eqnarray}
\label{ya}
I^a=&&{\sum_i}F_i^a,\nonumber\\
Y^a=&&\frac{\mu}{\mu+\nu}I_{1}^{a}+\frac{\nu}{\mu+\nu}I_{2}^{a}+\frac{i\lambda}{2(\mu+\nu)}
f_{abc}\sum_{i\neq j}{\omega}_{ij}I_i^{b}I_j^{c}\nonumber\\
&&(i,j=1,2).
\end{eqnarray}
Here $I^a$ form a $SU(3)$ algebra characterized by $f_{abc}$,
$\{F_{i}^{a},a=1,2,\cdots,8\}$ form a local $su(3)$ on the $i$ site
and they obey the commutation relation
\begin{eqnarray}
\label{ff} [F_{i}^{a},F_{j}^{b}]=i f_{abc}{\delta}_{ij}F_{i}^{c}.
\end{eqnarray}
$\mu$, $\nu$, $\lambda$ are parameters or Casimir operators and
$({\omega}_{ij}=-{\omega}_{ji})$ which satisfies \nonumber\\

$ \omega_{ij}=\left\{ \begin{array}{cc}
1, & {\rm if}\;\;\;\; i>j\\
-1, & {\rm if}\;\;\;\; i<j\\
0, & {\rm if}\;\;\;\; i=j
\end{array}
\right.$.\nonumber\\
Eq.(\ref{ya}) plays an important role in explaining the physical
meaning of the representation theory of Chari-Pressley \cite{3chari}
through more calculation \cite{3Bai}.

If we take the condition $\mu\nu$=-$\frac{{\lambda}^2}{4}$, in terms
of the notations
\begin{eqnarray}
\label{notation1} &&\tilde{I}^{\pm}={Y}^{1}{\pm}
iY^{2},\;\;\;\;{\tilde{{U}}^{\pm}}=Y^{6}{\pm} iY^{7},\;\;\;\;
{\tilde{{V}}^{\pm}}=Y^{4}{\pm} iY^{5},\nonumber\\
&&{\tilde{I}^{3}}=Y^{3},\;\;\;\;\;\;\;\;\;\;\;\;\;{\tilde{I}^{8}}=\frac{2}{\sqrt{3}}Y^{8},
\end{eqnarray}

we can directly calculate and obtain
\begin{eqnarray}
\label{y2} ({\bf{Y}})^2=\frac{1}{3}.
\end{eqnarray}

In the following we get the commutation relations
\begin{equation}
\label{breveI}
\begin{array}{l}
\begin{array}{lll}
\vspace{0.2cm}
[{\tilde{I}^{+}},{\tilde{I}^{-}}]=2{\tilde{I}^{3}}&\;\;
[{\tilde{{I}}^{3}},{\tilde{{I}}^{8}}]=[{\tilde{{I}}^{\pm}},{\tilde{{I}}^{8}}]=0

\cr \vspace{0.2cm}
[{\tilde{{I}}^{3}},{\tilde{I}^{\pm}}]={\pm}{\tilde{{I}}^{\pm}}&\;\;
[{\tilde{{I}}^{3}},{\tilde{U}^{\pm}}]={\mp}\frac{1}{2}{\tilde{U}^{\pm}}

\cr \vspace{0.2cm}
[\tilde{{I}}^{8},{\tilde{U}^{\pm}}]={\pm}{\tilde{U}^{\pm}}&\;\;
[{\tilde{I}^{3}},{\tilde{V}^{\pm}}]={\mp}\frac{1}{2}{\tilde{V}^{\pm}}

\cr \vspace{0.2cm}
[\tilde{I}^{8},\tilde{V}^{\pm}]={\mp}{\tilde{V}^{\pm}}&\;\;
[{\tilde{U}^{+}},{\tilde{U}^{-}}]=2\tilde{U}^3

\cr \vspace{0.2cm}
[{\tilde{I}^{\pm}},{\tilde{U}^{\pm}}]={\pm}{\tilde{V}^{\mp}}&\;\;
[{\tilde{V}^{+}},{\tilde{V}^{-}}]=2\tilde{V}^3

\cr \vspace{0.2cm}
[{\tilde{V}^{\pm}},{\tilde{I}^{\pm}}]={\pm}{\tilde{U}^{\mp}}&\;\;
[{\tilde{U}^{\pm}},{\tilde{V}^{\pm}}]={\pm}{\tilde{I}^{\mp}}

 \cr \vspace{0.2cm}
[{\tilde{I}^{\pm}},{\tilde{U}^{\pm}}]={\pm}{\tilde{V}^{\mp}}&\;\;
[{\tilde{V}^{\mp}},{\tilde{I}^{\pm}}]=[{\tilde{U}^{\pm}},{\tilde{V}^{\mp}}]=0

\cr \vspace{0.2cm} [{\tilde{I}^{\mp}},{\tilde{U}^{\pm}}]=0,
\end{array}
\end{array}
\end{equation}
where $\tilde{U}^3$ =
$-\frac{1}{2}\tilde{I}^{3}+\frac{3}{4}\tilde{I}^8$ and $
\tilde{V}^3$ = $-\frac{1}{2}\tilde{I}^{3}-\frac{3}{4}\tilde{I}^8$.
They are similar with the commutation relations of the $I$-spin,
$U$-spin, and $V$-spin.

Setting $x$ the eigenvalue of $\tilde{I}^3$, then
$|xE-\tilde{I}^3|=0$ with $E$ a unite matrix and its solutions are
\begin{eqnarray}
\label{x2}
&&x_1=0,\;\;\;\;x_{2,3}=\pm\frac{1}{2},\nonumber\\
&&x_{4,5}=\pm\frac{1}{2(\mu+\nu)}\sqrt{\mu^2-2\mu\nu+\nu^2-\lambda^2},\nonumber\\
&&x_{6,7}=\frac{1}{4}\pm\frac{1}{4(\mu+\nu)}\sqrt{\mu^2-2\mu\nu+\nu^2-\lambda^2}\nonumber\\
&&x_{8,9}=-\frac{1}{4}\pm\frac{1}{4(\mu+\nu)}\sqrt{\mu^2-2\mu\nu+\nu^2-\lambda^2}.
\end{eqnarray}
By taking $\mu\nu=-\frac{\lambda^2}{4}$, we can get
\begin{eqnarray}
\label{x}
x_{1,2,3}=0,\;\;\;\;x_{4,5,6}=\frac{1}{2},\;\;\;\;x_{7,8,9}=-\frac{1}{2}.
\end{eqnarray}
If we take the similar matrix as
\begin{eqnarray}
\label{su3a} A=\left(\matrix{1&0&0&0&0&0&0&0&0 \cr
0&\nu&0&-\frac{\lambda}{2}&0&0&0&0&0 \cr
0&0&\nu&0&0&0&-\frac{\lambda}{2}&0&0 \cr
0&-\frac{\lambda}{2}&0&\nu&0&0&0&0&0 \cr 0&0&0&0&1&0&0&0&0 \cr
0&0&0&0&0&\nu&0&-\frac{\lambda}{2}&0 \cr
0&0&-\frac{\lambda}{2}&0&0&0&\nu&0&0 \cr
0&0&0&0&0&-\frac{\lambda}{2}&0&\nu&0 \cr 0&0&0&0&0&0&0&0&1}\right),
\end{eqnarray}
and its inverse matrix $A^{-1}$ can be obtained easily, thus we can
obtain
\begin{equation}
\label{iuv38'}
\begin{array}{ll}
\vspace{0.4cm} \bar{I}^{3}=A^{-1}{\tilde{I}^{3}}A=\left(
\matrix{{I}^{3}&0&0 \cr 0&I^{3}&0 \cr 0&0&I^{3}}\right)&

\cr \vspace{0.4cm}

\bar{I}^{8}=A^{-1}{\tilde{I}^{8}}A=\left( \matrix{I^{8}&0&0 \cr
0&I^{8}&0 \cr 0&0&I^{8}}\right)

\cr \vspace{0.4cm}

\bar{I}^{+}=A^{-1}{\tilde{I}^{+}}A=\left( \matrix{{\alpha}I^{+}&0&0
\cr 0&{\alpha}^{-1}I^{+}&0 \cr 0&0&I^{+}}\right)&

\cr \vspace{0.4cm}

 \bar{I}^{-}=A^{-1}{\tilde{I}^{-}}A=\left(
\matrix{{\alpha}^{-}I^{-}&0&0 \cr 0&{\alpha}I^{-}&0 \cr
0&0&I^{-}}\right)

\cr  \vspace{0.4cm}

 \bar{U}^{+}=A^{-1}{\tilde{U}^{+}}A=\left(
\matrix{U^{+}&0&0 \cr 0&{\alpha}U^{+}&0 \cr
0&0&{\alpha}^{-}U^{+}}\right)&

\cr \vspace{0.4cm}

\bar{U}^{-}=A^{-1}{\tilde{U}^{-}}A=\left( \matrix{ U^{-}&0&0 \cr
0&{\alpha}^{-1}U^{-}&0 \cr 0&0&{\alpha}U^{-}}\right)

\cr  \vspace{0.4cm}

\bar{V}^{+}=A^{-1}{\tilde{V}^{+}}A=\left(
\matrix{{\alpha}^{-}V^{+}&0&0 \cr 0&V^{+}&0 \cr
0&0&{\alpha}V^{+}}\right)&

\cr \vspace{0.4cm}

\bar{U}^{-}=A^{-1}{\tilde{V}^{-}}A=\left( \matrix{ {\alpha}V^{-}&0&0
\cr 0&V^{-}&0 \cr 0&0&{\alpha}^{-1}V^{-}}\right)

\end{array}
\end{equation}
where $\alpha=\nu-\frac{\lambda}{2}$. In virtue of a similar
transformation, we can reduce the eight $9\times9$ matrix to the
three $3\times3$ block diagonal, so it is marked as the reduced
$Y(su(3))$ algebra in this case. Taking the correspondence
\begin{eqnarray}
\label{y'} &&I^{+}{ \rightarrow}{\alpha}I^{+},\;\;\;I^{-}{
\rightarrow}{\alpha}^{-1}I^{-},\;\;\;
I^{3}{ \rightarrow}I^{3},\;\;\;I^{8}{ \rightarrow}I^{8},\nonumber\\
&&U^{+}{ \rightarrow}{\alpha}U^{+},\;\;U^{-}{
\rightarrow}{\alpha}^{-1}U^{-},\nonumber\\&& V^{+}{
\rightarrow}{\alpha}V^{+},\;\;V^{-}{
\rightarrow}{\alpha}^{-1}V^{-}.\nonumber
\end{eqnarray}
Eq. (\ref{breveI}) will get the same result, that is, the Yangian
algebra we discussed hides a $u(1)$ algebra.

\section{The applications of $Y(su(3))$ Algebra in the mixed light pseudoscalar meson state $\eta$}

The $\eta$ and $\eta^{'}$ mesons play the important role in low
energy QCD. $\eta-\eta^{'}$ mixing system is one of the most
attractive problems all along \cite{Tian1,bbr,hfj,rej}. Hence we
choose the superposition of singlet and octet of $su(3)$ as the
initial state
\begin{eqnarray}
|\eta\rangle=\alpha_1|\eta^{0'}\rangle+\alpha_2|\eta^0\rangle,
\end{eqnarray}
where $\alpha_1$ and $\alpha_2$ are the normalized real amplitudes
and satisfy $\alpha_1^2+\alpha_2^2=1$.
$|\eta^{0'}\rangle=|\Omega^0\rangle=\frac{1}{\sqrt{3}}(|u\bar{u}\rangle+|d\bar{d}\rangle+|s\bar{s}\rangle)$,
$|\eta^{0}\rangle=\frac{1}{\sqrt{6}}(-|u\bar{u}\rangle-|d\bar{d}\rangle+2|s\bar{s}\rangle)$.

As is known, the entanglement degree of the genuine N-particle
qutrit pure state \cite{Pan} is measured by its mean entropy
\cite{shw}
\begin{eqnarray}
\label{2} C^{(N)}_\Phi=\left\{
\begin{array}{l}
\frac1N\sum_{i=1}^N
S_{(i)}\;\;\;\;\;$if$\;S_i\neq0\;\forall\;i\\
0\;\;\;\;\;\;\;\;\;\;\;\;\;\;\;\;\;\;\;\;\;\;$otherwise$
\end{array} \right.,
\end{eqnarray}
where $S_i=-Tr((\rho_\Phi)_iLog_3(\rho_\Phi)_i)$ is the reduced
partial Von Neumann entropy for the $i$th particle only, with the
other $N-1$ particles traced out, and $(\rho_\Phi)_i$ is the
corresponding reduced density matrix. The system we discuss here is
bipartite qutrit, $N=2$. Thus the entanglement degree of the initial
state $|\eta\rangle$ can be gotten as
\begin{eqnarray}
C_{ini}=-2(\frac{\alpha_1}{\sqrt{3}}-\frac{\alpha_2}{\sqrt{6}})^2Log_3(\frac{\alpha_1}{\sqrt{3}}-\frac{\alpha_2}{\sqrt{6}})^2\nonumber\\
-(\frac{\alpha_1}{\sqrt{3}}+\frac{2\alpha_2}{\sqrt{6}})^2Log_3(\frac{\alpha_1}{\sqrt{3}}+\frac{2\alpha_2}{\sqrt{6}})^2.
\end{eqnarray}
The value of $C_{ini}$ has been discussed in detail with range from $0$ to $1$ \cite{Tian1}.

To illustrate the effects of every generator of Yangian $Y(su(3))$
on the initial state, we will take the following operators as the
transition operators in the general and reduced Yangian cases
respectively.
\begin{table}[t]
\begin{center}
\caption{The general Yangian in decay channels.}
\begin{tabular}{|c||l|c|l|l|} \hline

$|\eta\rangle_{ini}$ & P & $|\eta\rangle_{fin}$ & $C_{fin}$ & decay  \\ [0.5ex] \hline
 $|\eta\rangle=\alpha_1|\eta^{0'}\rangle+\alpha_2|\eta^0\rangle$¡¡& $\tilde{I}^{+}$ & $|\eta_1\rangle$ & $C_1$ & $\eta\rightarrow\pi^{+}\pi^{-}$ \\
¡¡¡¡¡¡¡¡& $\tilde{I}^{-}$ & $|\eta_2\rangle$  & $C_1$ & $\eta\rightarrow\pi^{+}\pi^{-}$ \\
¡¡& $\tilde{U}^{+}$ &$|\eta_3\rangle$  &$C_1$ &$\eta\rightarrow K^{0}\bar{K}^{0}$ \\
 ¡¡¡¡¡¡ & $\tilde{U}^{-}$ & $|\eta_4\rangle$  & $C_1$ & $\eta\rightarrow K^{0}\bar{K}^{0}$  \\
 ¡¡¡¡¡¡ & $\tilde{V}^{+}$ & $|\eta_5\rangle$  & $C_1$ & $\eta\rightarrow K^{+}K^{-}$  \\
 ¡¡ & $\tilde{V}^{-}$ & $|\eta_6\rangle$  & $C_1$ & $\eta\rightarrow K^{+}K^{-}$  \\
 ¡¡ & $\tilde{I}^{3}$ & $|\eta_7\rangle$  & $Log_3 2$ & $\eta\rightarrow \pi^{0}$  \\
 ¡¡ & $\tilde{I}^{8}$ & $|\eta_8\rangle$  & $Log_3 2$ & $\eta\rightarrow \eta^{0}\eta^{0'}$ \\
 \hline
\end{tabular}
\end{center}
\end{table}

(i) The general $Y(su(3))$

Every general transitions operator of $Y(su(3))$ is applied to the initial state, then we can obtain the following the results.
\begin{eqnarray}
&&|\eta_1\rangle=\tilde{I}^{+}|\eta\rangle=(\frac{\alpha_1}{\sqrt{3}}-\frac{\alpha_2}{\sqrt{6}})\frac{1}{\mu+\nu }\nonumber\\&&\;\;\;\;\;\;\;[(\mu-\frac{\lambda}{2})|u\bar{d}\rangle+(\nu+\frac{\lambda}{2})|d \bar{u}\rangle],\nonumber\\
&&|\eta_2\rangle=\tilde{I}^{-}|\eta\rangle=(\frac{\alpha_1}{\sqrt{3}}-\frac{\alpha_2}{\sqrt{6}})\frac{1}{\mu+\nu }\nonumber\\&&\;\;\;\;\;\;\;[(\nu+\frac{\lambda}{2})|u\bar{d}\rangle+(\mu-\frac{\lambda}{2})|d \bar{u}\rangle],\nonumber\\
&&|\eta_3\rangle=\tilde{U}^{+}|\eta\rangle=(\frac{\alpha_1}{\sqrt{3}}+\frac{2\alpha_2}{\sqrt{6}})\frac{1}{\mu+\nu }\nonumber\\&&\;\;\;\;\;\;\;[(\mu-\frac{\lambda}{2})|d\bar{s}\rangle+(\nu+\frac{\lambda}{2})|s \bar{d}\rangle],\nonumber\\
&&|\eta_4\rangle=\tilde{U}^{-}|\eta\rangle=(\frac{\alpha_1}{\sqrt{3}}-\frac{\alpha_2}{\sqrt{6}})\frac{1}{\mu+\nu }\nonumber\\&&\;\;\;\;\;\;\;[(\nu+\frac{\lambda}{2})|d\bar{s}\rangle+(\mu-\frac{\lambda}{2})|s \bar{d}\rangle],\nonumber\\
&&|\eta_5\rangle=\tilde{V}^{+}|\eta\rangle=(\frac{\alpha_1}{\sqrt{3}}-\frac{\alpha_2}{\sqrt{6}})\frac{1}{\mu+\nu }\nonumber\\&&\;\;\;\;\;\;\;[(\nu+\frac{\lambda}{2})|u\bar{s}\rangle+(\mu-\frac{\lambda}{2})|s \bar{u}\rangle],\nonumber\\
&&|\eta_6\rangle=\tilde{V}^{-}|\eta\rangle=(\frac{\alpha_1}{\sqrt{3}}+\frac{2\alpha_2}{\sqrt{6}})\frac{1}{\mu+\nu }\nonumber\\&&\;\;\;\;\;\;\;[(\mu-\frac{\lambda}{2})|u\bar{s}\rangle+(\nu+\frac{\lambda}{2})|s \bar{u}\rangle],\nonumber\\
&&|\eta_7\rangle=\tilde{I}^{3}|\eta\rangle=(\frac{\alpha_1}{\sqrt{3}}-\frac{\alpha_2}{\sqrt{6}})\frac{1}{2}(|u\bar{u}\rangle-|d \bar{d}\rangle),\nonumber\\
&&|\eta_8\rangle=\tilde{I}^{8}|\eta\rangle=(\frac{\alpha_1}{\sqrt{3}}-\frac{\alpha_2}{\sqrt{6}})\frac{1 }{3}(|u\bar{u}\rangle+|d \bar{d}\rangle)\nonumber\\&&\;\;\;\;\;\;\;-(\frac{\alpha_1}{\sqrt{3}}+\frac{2\alpha_2}{\sqrt{6}})\frac{2}{3}|s\bar{s}\rangle
\end{eqnarray}
The results mentioned above are summarized in Table I. We can note
that the entanglement degrees of the finial states are the same
after the actions of the operators
$\tilde{I}^{\pm}$,$\tilde{U}^{\pm}$ and $\tilde{V}^{\pm}$. $C_1=
-(\mu-\frac{\lambda}{2})^{2} Log_3 (\mu-\frac{\lambda}{2})^{2}
-(\nu+\frac{\lambda}{2})^{2} Log_3 (\nu+\frac{\lambda}{2})^{2}$,
with $(\mu-\frac{\lambda}{2})^{2}+(\nu+\frac{\lambda}{2})^2=1$.
As we can see, by changing the value of $\mu$ and $\lambda$, the entanglement degree C
can be tuned (due to the normalization condition, $\nu$ is not an independent parameter).
The behavior of $C_1$ depends on $\mu$ for the case of $\lambda=2$, as shown in Fig. 1.
It is worth noting that the maximum value of $C_1$ is $Log_3 2$.
In addition, we can find that the entanglement degrees of the finial
states are same and the maximum value of $C_1$ after the actions of the operators
$\tilde{I}^{3}$ and $\tilde{I}^{8}$.

\begin{figure}[h]
\includegraphics[angle=0,width=8cm]{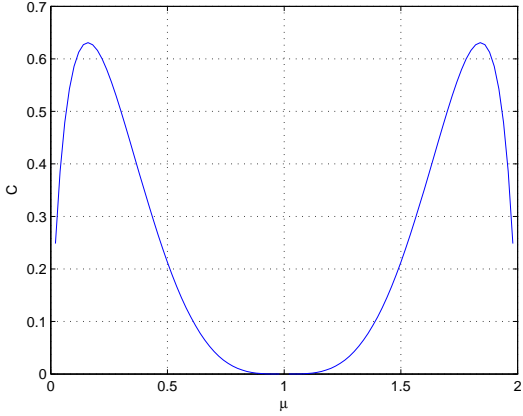}
\caption{(Color online) The evolution of the entanglement degree $C_1$ varies with $\mu$, given $\lambda=2$.}
\end{figure}

(ii) The reduced $Y(su(3))$

\begin{table}[t]
\begin{center}
\caption{The reduced Yangian in decay channels.}
\begin{tabular}{|c||l|c|l|l|} \hline

$|\eta\rangle_{ini}$ & $P^{'}$ & $|\eta^{'}\rangle_{fin}$ & $C^{'}_{fin}$ & decay  \\ [0.5ex] \hline
 $|\eta\rangle=\alpha_1|\eta^{0'}\rangle+\alpha_2|\eta^0\rangle$¡¡& $\bar{I}^{+}$ & $|\eta_1^{'}\rangle$ & $0$ & $\eta\rightarrow\pi^{-}$  \\
¡¡¡¡¡¡¡¡& $\bar{I}^{-}$ & $|\eta_2^{'}\rangle$  & $0$ & $\eta\rightarrow\pi^{+}$  \\
¡¡& $\bar{U}^{+}$ &$|\eta_3^{'}\rangle$  &$0$ &$\eta\rightarrow \bar{K}^{0}$  \\
 ¡¡¡¡¡¡ & $\bar{U}^{-}$ & $|\eta_4^{'}\rangle$  & $0$ & $\eta\rightarrow K^{0}$   \\
 ¡¡¡¡¡¡ & $\bar{V}^{+}$ & $|\eta_5^{'}\rangle$  & $0$ & $\eta\rightarrow K^{+}$ \\
 ¡¡ & $\bar{V}^{-}$ & $|\eta_6^{'}\rangle$  & $0$ & $\eta\rightarrow K^{-}$  \\
 ¡¡ & $\bar{I}^{3}$ & $|\eta_7^{'}\rangle$  & $Log_3 2$ & $\eta\rightarrow \pi^{0}$   \\
 ¡¡ & $\bar{I}^{8}$ & $|\eta_8^{'}\rangle$  & $Log_3 2$ & $\eta\rightarrow \eta^{0}\eta^{0'}\pi^{0}$   \\
 \hline
\end{tabular}
\end{center}
\end{table}
Every reduced transitions operator of $Y(su(3))$ is applied to the initial state by the same way, then we can obtain the following the results.
\begin{eqnarray}
&&|\eta_1^{'}\rangle=\bar{I}^{+}|\eta\rangle=\xi^{-1}(\frac{\alpha_1}{\sqrt{3}}-\frac{\alpha_2}{\sqrt{6}})|d \bar{u}\rangle\nonumber\\&&\;\;\;\;\;\;\;=\xi^{-1}(\frac{\alpha_1}{\sqrt{3}}-\frac{\alpha_2}{\sqrt{6}})|\pi^{-}\rangle\nonumber\\
&&|\eta_2^{'}\rangle=\bar{I}^{-}|\eta\rangle=\xi^{-1}(\frac{\alpha_1}{\sqrt{3}}-\frac{\alpha_2}{\sqrt{6}})|u \bar{d}\rangle\nonumber\\&&\;\;\;\;\;\;\;=-\xi^{-1}(\frac{\alpha_1}{\sqrt{3}}-\frac{\alpha_2}{\sqrt{6}})|\pi^{+}\rangle\nonumber\\
&&|\eta_3^{'}\rangle=\bar{U}^{+}|\eta\rangle=\xi^{-1}(\frac{\alpha_1}{\sqrt{3}}+\frac{2\alpha_2}{\sqrt{6}})|s \bar{d}\rangle\nonumber\\&&\;\;\;\;\;\;\;=\xi^{-1}(\frac{\alpha_1}{\sqrt{3}}+\frac{2\alpha_2}{\sqrt{6}})|\bar{K}^{0}\rangle\nonumber\\
&&|\eta_4^{'}\rangle=\bar{U}^{-}|\eta\rangle=\xi^{-1}(\frac{\alpha_1}{\sqrt{3}}-\frac{\alpha_2}{\sqrt{6}})|d \bar{s}\rangle\nonumber\\&&\;\;\;\;\;\;\;=\xi^{-1}(\frac{\alpha_1}{\sqrt{3}}-\frac{\alpha_2}{\sqrt{6}})|K^{0}\rangle\nonumber\\
&&|\eta_5^{'}\rangle=\bar{V}^{+}|\eta\rangle=\xi^{-1}(\frac{\alpha_1}{\sqrt{3}}-\frac{\alpha_2}{\sqrt{6}})|u \bar{s}\rangle\nonumber\\&&\;\;\;\;\;\;\;=\xi^{-1}(\frac{\alpha_1}{\sqrt{3}}-\frac{\alpha_2}{\sqrt{6}})|K^{+}\rangle\nonumber\\
&&|\eta_6^{'}\rangle=\bar{V}^{-}|\eta\rangle=\xi^{-1}(\frac{\alpha_1}{\sqrt{3}}+\frac{2\alpha_2}{\sqrt{6}})|s \bar{u}\rangle\nonumber\\&&\;\;\;\;\;\;\;=\xi^{-1}(\frac{\alpha_1}{\sqrt{3}}+\frac{2\alpha_2}{\sqrt{6}})|K^{-}\rangle\nonumber\\
&&|\eta_7^{'}\rangle=\bar{I}^{3}|\eta\rangle=(\frac{\alpha_1}{\sqrt{3}}-\frac{\alpha_2}{\sqrt{6}})\frac{1}{2}(|u\bar{u}\rangle-|d \bar{d}\rangle) \nonumber\\&&\;\;\;\;\;\;\;=(\frac{\alpha_1}{\sqrt{3}}-\frac{\alpha_2}{\sqrt{6}})\frac{1}{\sqrt{2}}|\pi^{0}\rangle\nonumber\\
&&|\eta_8^{'}\rangle=\bar{I}^{8}|\eta\rangle=(\frac{\alpha_1}{\sqrt{3}}-\frac{\alpha_2}{\sqrt{6}})(\frac{1}{3}|u \bar{u}\rangle-\frac{1}{2}|d \bar{d}\rangle)\nonumber\\&&\;\;\;\;\;\;\;-\frac{2}{3}(\frac{\alpha_1}{\sqrt{3}}+\frac{2\alpha_2}{\sqrt{6}})|s\bar{s}\rangle\nonumber\\
&&\;\;\;\;\;\;\;=(-\frac{5\alpha_1}{18}-\frac{7\alpha_2}{18\sqrt{2}})|\eta^{0'}\rangle-(\frac{7\alpha_1}{18\sqrt{2}}+\frac{17\alpha_2}{36})|\eta^{0}\rangle
\nonumber\\&&\;\;\;\;\;\;\;+(\frac{5\alpha_1}{6\sqrt{6}}-\frac{5\alpha_2}{12\sqrt{3}})|\pi^{0}\rangle
\end{eqnarray}

These results are summarized in Table II. Table II can show the
effects of the reduced Yangian generators. From the above Table II,
to our surprise, we find that the reduced Yangian  generators
($\bar{I}^{\pm}$,$\bar{U}^{\pm}$ and $\bar{V}^{\pm}$) make the
initial state disentangled, however, $\bar{I}^{3}$ and $\bar{I}^{8}$
can not change the entanglement degree of the initial state in the
case of the normalization condition. By comparing the effects of the
general and reduced Yangian generators based on their transition
effect on the entanglement of the initial state, we can find easily
the difference of the transition effect between the general and
reduced Yangian generators by Tables I and II.

\section{Conclusion}

In this paper, we have studied the applications of Yangian
$Y(sl(2))$ and $Y(su(3))$ for quantum entanglement in the general
and reduced Yangian cases. For $Y(sl(2))$, we only studied the
effects Yangian algebra on the entanglement. By calculating, we can
find that the general transition operators Yangian $Y(sl(2))$ can
not make the initial state disentangled, and the entanglement degree
of the finial state is from $0$ to $1$. However, the reduced ones
can make the initial state disentangled except $Y^3$. It is worth
noting that the general transition operators, $J^3$ and $Y^3$ can
not change the entanglement of the initial state.

For $Y(su(3))$, we have studied not only the change of entanglement
degree, but also the decay channel in the mixed light pseudoscalar
meson states. Our results show that the general $Y(su(3))$
generators can not make the initial state disentangled, but the
reduced ones can make the initial state disentangled except
$\bar{I}^{3}$ and $\bar{I}^{8}$. In addition, we can obtain that
$\tilde{I}^{3}$, $\tilde{I}^{8}$,$\bar{I}^{3}$ and $\bar{I}^{8}$ can
not change the entanglement of the initial state, and the
entanglement degree of the initial and final states are $Log_3 2$
with the normalization condition. Moreover, some hadronic decay
channels of pseudo-scalar mesons can be reformulated under the
framework of Yangian by acting transition operators consisting of
generators of Yangian $Y (su(3))$ on the initial state.

Now a question put forward: can we use this method to $su(n)$? To
our knowledge, this problem has not been discussed in this thesis.
But we can surmise that for $Y(su(n))$ the matrices of the
generators ${\bf{Y}}$ can be written as $n$ pieces of $n$ $\times$
$n$ matrices, furthermore each pieces is formed by the consequently
generators of $su(n)$. Consequently, it is of interesting area how
to generalize the idea in $Y(su(n))$, which makes the system contact
with physical application.

\section{Acknowledgements} This work is in part supported by the NSF of
China under Grant No. 11075101, Shanghai Leading Academic Discipline
Project (Project number S30105). L.G.Qin is also partially supported
by the Ph.D. Program Foundation of Ministry of Education of China
under Grant No. 20093108110004. The authors are grateful to Xin-Jian
Xu for valuable discussions.

\end{document}